\documentclass[amssymb,amsmath,aps,12pt,superscriptaddress,longbibliography]{revtex4-1}

\usepackage{graphicx,mdwlist,color,wrapfig,revsymb}
\usepackage{etoolbox}

\begin{document}

\title{Theory of topological excitations and metal-insulator transition in reentrant integer quantum Hall effect}

\author{George Simion}
\affiliation{Department of Physics and Astronomy, Purdue University, West Lafayette, Indiana 47907 USA}
\author{Tsu-ging Lin}
\affiliation{Department of Physics and Astronomy, Purdue University, West Lafayette, Indiana 47907 USA}
\author{John D. Watson}
\affiliation{Department of Physics  and Astronomy, Purdue University, West Lafayette, Indiana 47907 USA}
\affiliation{Birck Nanotechnology Center, Purdue University, West Lafayette, Indiana 47907 USA}
\author{Michael J. Manfra}
\affiliation{Department of Physics  and Astronomy, Purdue University, West Lafayette, Indiana 47907 USA}
\affiliation{Birck Nanotechnology Center, Purdue University, West Lafayette, Indiana 47907 USA}
\affiliation{School of Electrical and Computer Engineering, Purdue University, West Lafayette, Indiana 47907 USA}
\affiliation{School of Materials Engineering, Purdue University,
West Lafayette, Indiana 47907 USA}
\author{Gabor A. Cs\'athy}
\affiliation{Department of Physics  and Astronomy, Purdue University, West Lafayette, Indiana
47907 USA}
\affiliation{Birck Nanotechnology Center, Purdue University, West
Lafayette, Indiana 47907 USA}
\author{Leonid P. Rokhinson}
\affiliation{Department of Physics and Astronomy, Purdue University, West Lafayette, Indiana
47907 USA}\affiliation{Birck Nanotechnology Center, Purdue University, West
Lafayette, Indiana 47907 USA} \affiliation{School of Electrical and Computer Engineering, Purdue University,
West Lafayette, Indiana 47907 USA}
\author{Yuli Lyanda-Geller}
\email{yuli@purdue.edu} \affiliation{Department of Physics and Astronomy, Purdue University, West Lafayette, Indiana
47907 USA} \affiliation{Birck Nanotechnology Center, Purdue University, West
Lafayette, Indiana 47907 USA}

\date{11 March 2017}

\begin{abstract}
The reentrant integer quantum Hall effects (RIQHE) are due to formation of electronic crystals. We show analytically and numerically that topological textures in the charge density distribution in these crystals in the vicinity of charged defects strongly reduce energy required for current-carrying excitations. 
The theory quantitatively explains sharp insulator-metal transitions experimentally observed in RIQHE states. The insulator to metal transition in RIQHE emerges as a thermodynamic unbinding transition of topological charged defects.
\end{abstract}

\maketitle

Topology and symmetry define states of matter and their response to external forces. Topological excitations dramatically alter the responses, but are difficult to predict because they cannot be obtained perturbatively. Here we find novel topological excitations of two-dimensional (2D) electrons in a perpendicular magnetic field $\vec{B} $.  The spectrum of electrons in this system is given by Landau levels, and interactions cause a variety of ground states as a function of $\vec{B} $. When a filling factor $\nu=\Phi_0/\Phi$ is integer or a certain fraction ($\Phi_0=h/e$ is a flux quanta, $\Phi=B/n$ is magnetic flux per electron, and $n$ is the electron density), the Hall resistance is quantized and longitudinal resistance vanishes, the hallmarks of the integer and fractional quantum Hall effect\cite{Klitzing1980,tsui82}. 

Besides the large family of fractional quantum Hall states, the electronic solid phases form a second distinct class of ground states for the 2DEG. These electronic solids break the translational and rotational symmetries to various degrees. The most well-known of the electronic solids is the Wigner solid at large magnetic fields\cite{Wigner1934}. However, 2DEGs also exhibit other electronic solids.
At $\nu>4$ charge density wave states arise: the unidirectional stripe phase is formed near a half-integer $\nu$, while bubble phases in certain ranges of fractional $\nu$ lead to the Hall resistance quantized to the nearest integer, i.e., to the reentrant integer quantum Hall effect (RIQHE)\cite{koulakov96,fogler96,moessner96,du99,lilly99,Eisenstein2001,Eisenstein2001a}. 
Bubble phases are insulating: longitudinal resistance vanishes at low temperatures. At a higher temperatures a non-zero resistivity emerges. The dc and microwave transport response\cite{Cooper1999,Eisenstein2002,Lewis2002,Wang2015,Msall2015} and the temperature dependence\cite{Deng2012a,Deng2012b} of the bubble phases have been under intense investigation.  However the nature of the observed sharp metal-insulator transition and the physical origin of excitations are not known in these systems.

In this Letter we propose a theory of metal to insulator transition and uncover topological origin of excitations in bubble phases, explaining an experimentally studied 
magnetic field-temperature phase diagram of the bubble phase.
The critical question is the physical mechanism of the electron transport. The ground RIQHE state is believed to be a crystal of bubbles carrying integer number of electrons  \cite{fogler96,Fogler1997}. Electron hopping between bubbles is forbidden by the Coulomb blockade, yielding an insulating state \cite{Eisenstein2001,Eisenstein2001a,fogler-review,Goerbig2003,Cote2003,Ettouhami2006}. Experiments show that for a given bubble phase, the metal-insulator transition temperature is the highest for the filling factor at the center of the range of magnetic fields characterizing the bubble crystal and is smaller on the the flanks of this range. These results preclude an interpretation of the metal to insulator transition as a consequence of melting of the bubble crystal as a whole due to  dislocations  \cite{Thouless1978,Fisher1979}. In such case the transition temperature would behave inversely proportional to the lattice constant of the bubble crystal, i.e., monotonically from one flank of the bubble phase range to the other, with no maximum of the transition temperature at the center. Thus, a different physics is involved here.

What may provide a conduction mechanism are charge defects in a bubble crystal (an extra electron or lack of an electron on a bubble).  Here we show that  in order to lower the energy cost of charge defects, crystalline bubbles around them acquire elongated dumbbell shape and form topological textures with vortex or 2D hedgehog symmetry, depending on the defect charge.  While topological textures of charged objects arise in bilayer electron systems \cite{Moon1995}, this phenomenon is unique and unanticipated in single-layer systems. At low defect densities, controlled by temperature and magnetic field, textures do not overlap and form an insulating crystal, similar to the Wigner or Abrikosov lattice \cite{Abrikosov1957,Wigner1934}. At high defect density topological defects  overlap, and their interactions are described by the XY-model. Because this occurs at temperatures above the Berezinski-Kosterlitz-Thouless phase transition \cite{Berezinskii1972,Kosterlitz1973}, the crystal of topological defects melts resulting in a sharp insulator-metal transition. This new phase transition resembles asymptotic freedom of quarks requiring them to be "squeezed" in order to be freed\cite{Gross1973}. We show that heterostructure disorder modifies bubble crystals and creates charge-neutral textures in the ground state, which affect melting temperature of the crystal of charged topological excitations, and metal-insulator transition.

The charge density wave phases at partial $\nu$ are conventionally described via the Hartree-Fock (HF) method \cite{fogler96,Fogler1997,aleiner95,fukuyama79}. We use
its ``interacting guiding centers" version \cite{Ettouhami2006} to study defects in two-electron bubble crystal corresponding to $\nu$ in our experiments here and in \cite{Deng2012b}.  The HF Hamiltonian is $H_{\rm{HF}}=1/2(\sum_i U_i(0)+ \sum_{i\neq j}U({\bf R}_{ij}))$ where indices $i$ and $j$ label lattice nodes, $U_i(0)$ is a charging energy required to put an extra electron on the two-electron bubble $i$, $\mathbf{R}_i$ is the coordinate of the bubble $i$, ${\bf R}_{ij}={\bf R}_{i}-{\bf R}_{j}$. The interaction energy $U({\bf R}_{ij})$ between bubbles $i$ and $j$ is given by
\begin{equation}
\label{eq:Ueff_gen}
 U({\bf{R}}_{ij})=\int{\frac{d\bf{q}}{4\pi^2} \rho_{i}^*({\bf {q}}) [V_{\rm{H}}({\bf{q}})-V_{\rm{F}}(\bf{q})]} \rho_{j}({\bf {q}})e^{i\mathbf{q}\cdot{\bf R}_{ij}},
\end{equation}
where $\rho_{i}$ is the site $i$ bubble density projected on the uppermost Landau Level (LL). The Hartree and exchange potentials are, respectively \cite{aleiner95}, $V_{\rm{H}}({\bf{q}})=\frac{2 \pi}{q}e^{-q^2/2}[L_{n}(q^2/2)]^2$, and $V_{\rm{F}}({\bf{q}})=2\pi\int{\frac{d^2\bf{q'}}{(2\pi)^2}V_{{\rm{H}}}({\bf{q}}) e^{-i ({\bf{q}}\times {\bf{q'}})\cdot\hat z}}$, where $L_{n}$ is the $n^{\rm{th}}$ Laguerre polynomial. The ground state of the system is an ideal bubble crystal with a triangular lattice of round bubbles. Creating single-bubble charge defects in an otherwise unperturbed bubble crystal costs $\sim 50K$, so bubbles around the defect must re-arrange themselves to lower the energy. One mechanism of re-arrangement is displacement of bubbles from lattice sites similar to displacement of electrons in a Wigner crystal due to vacancies \cite{Fisher1979}. Calculated energies of 1\={e} and 3\={e} defects in a crystal of round bubbles with account of such displacements $\sim 10K$, dashed lines in Fig.~\ref{f-defect-th}a.  However, such defects cannot explain high conductivity at 100 mK.

\begin{figure}
\def\ffile{f-defect-th}
\includegraphics[width=0.48\textwidth]{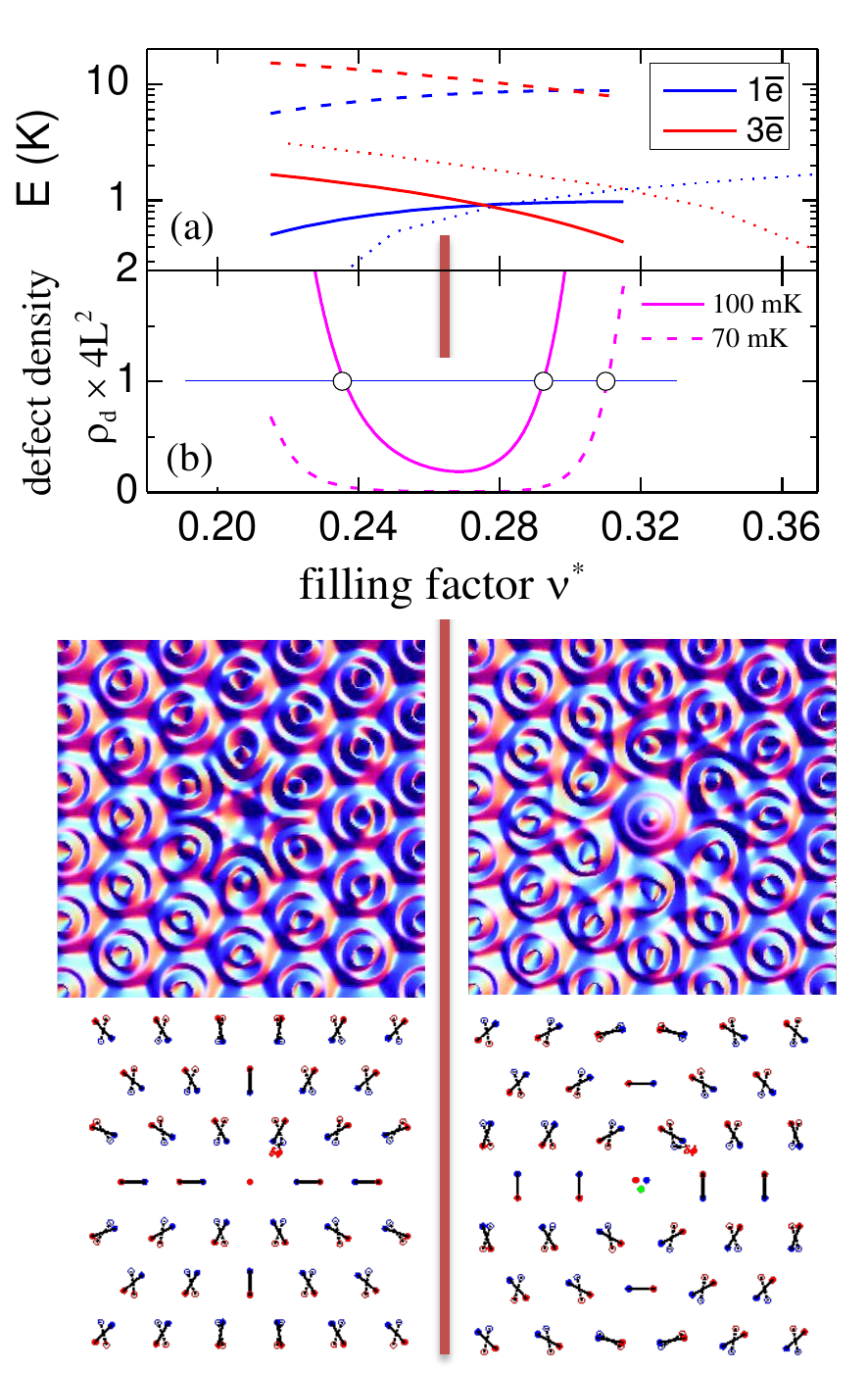}
\caption{(a) Activation energy for an isolated defect is calculated for round bubbles (Eq.~\ref{eq:2elebubble_int_S}, dashed lines) and textured defects using analytical theory (Eq.~\ref{eq:effH_pm_d}, solid lines) or full numerical calculations (dotted lines). Note that topological deformations reduce activation energies by a factor of 10. (b) Density of defects $\rho_d$ is calculated for 70 mK and 100 mK. The melting phase transition temperature $T_L$ corresponds to the defect density where defect separation $\approx 2L$. }
\label{\ffile}
\end{figure}

The other mechanism of rearrangement of bubble crystal that lowers energy of charged defects is a change of shape and elongation of two-electron bubbles around 1\={e} and 3\={e} defects. In the original proposal of the bubble state \cite{koulakov96,fogler96,Fogler1997}, electron guiding centers were on top of each other. However, the two electron states defining bubbles are not identical:  in the symmetric gauge they are given by 3LL wavefunctions $u_{-2}$ and $u_{-1}$ with angular momentum projections $m=-2$ and $m=-1$ \cite{Giuliani-book}. Defects displace charges in such states  differently, resulting in bubble elongations. Furthermore, when uniform positive background is included, such charge redistribution leads to dipole moments of bubbles. An elongated 2\=e bubble looks like a dumbbell with two charges (electron guiding centers) on it.
The separation between two dumbbell weights (electrons) appears to be smaller than the magnetic length $\lambda$.  This allows us to develop the method of solution for the problem of interacting electrons using the small parameter $\lambda/w$, $w$ is the bubble lattice constant.
In the experimentally relevant range of $\nu$, $w\sim  8\lambda$. We suggest the variational wavefunction of an elongated two-electron bubble with guiding
centers separated by $2\mathbf{a}$:
\begin{equation}
\label{eq:wf_2e_bubble_gen}
\Psi_{\xi,\bf{a}}({\bf{r}}_1,{\bf{r}}_2)=\alpha \left(\psi_{\bf{a}}({\bf{r}}_1,{\bf{r}}_2)
+\xi\psi_{-\bf{a}}({\bf{r}}_1,{\bf{r}}_2)\right)
\end{equation}
where $\alpha$ is the normalization coefficient, $\xi$ is a variational parameter, and
\begin{eqnarray*}
\psi_{\bf{a}}({\bf{r}}_1,{\bf{r}}_2)=[u_{-2}({\bf{r}}_1+{\bf{a}})
u_{-1}({\bf{r}}_2-{\bf{a}})e^{i({\bf{r}}_1-{\bf{r}}_2)\times
{\bf{a}}/2}\\ \nonumber
-u_{-1}({\bf{r}}_1-{\bf{a}})
u_{-2}({\bf{r}}_2+{\bf{a}})e^{i({\bf{r}}_2-{\bf{r}}_1)\times{\bf{a}}/2}].
\end{eqnarray*}
 This wavefunction is a superposition of two dumbbells with opposite orientation of weights corresponding to two  $m=-2$ and $m=-1$ electrons. It is used to find electron density of an elongated bubble projected on the 3LL and interaction energy between dumbbells.
The total energy includes a contribution from interactions between dumbbells and the charge defects. The wavefunction of a 1\={e} bubble $u_{-2}$  has round shape. For a 3\={e} bubble the wavefunction is a Slater determinant of $u_{-2}$, $u_{-1}$, and $u_{0}$. Its exact shape can be determined by energy minimization, but the physics is primarily determined by dumbbells, so we neglect the detailed structure of the 3\={e} bubble and model it as having all three guiding centers on top of each other.
Interactions of a single charge defect at a site $k$ with surrounding dumbbells labeled $i$ can be expressed in terms of vectors $\mathbf{\boldsymbol{\mu}_i}$, which are rescaled and rotated bubble elongations $\bf{a}_i$. Retaining terms up to $R_{ik}^{-3}$, we get  asymptotic expansion in $\lambda/w$ for the elongation-dependent contribution to energy:
\begin{eqnarray}
\label{eq:effH_pm_d}
 H^{\pm}&=&\sum_{i\neq
k}
\left[
\mp\left[\frac{\boldsymbol\mu_i\cdot \hat {\mathbf{R}}_{ik}}{R_{ik}^2}+
\frac{v_i}{2 R_{ik}^3} \cos{(2\varphi_i)}\right]+ U_f^{\pm}(i)\right]
\nonumber\\
&+&\frac{1}{2}\sum_{i \neq j}\frac{ {\boldsymbol{\mu}}_i \cdot {\boldsymbol{\mu}}_j- 3
({\boldsymbol{\mu}}_i \cdot \hat{{\mathbf{R}}}_{ij})({\boldsymbol{\mu}}_j \cdot
\hat{{\mathbf{R}}}_{ij}) } {R_{ij}^3}~,
\end{eqnarray}
where $(+)$ and $(-)$ signs correspond to 1\={e} and 3\={e} defects, $\hat{{\mathbf{R}}}_{ij}=\mathbf{R_{ij}}/R_{ij}$ and $\varphi_i$ is the angle between $\bf{R}_{ki}$ and $\bf{a}_i$. The first three terms of Eq. (\ref{eq:effH_pm_d}) come from the interaction of $i$-th dumbbell with the charge defect, the fourth term is a dipole-dipole interaction between dumbbells at sites $i$ and $j$. Analytical expressions for projected density, quantities $\mathbf{\mu_i}$, $v_i$,  and  $U_f^{\pm}(i)$ are given in supplementary material.
We next analytically minimize energy functional (\ref{eq:effH_pm_d}). Charge density in the vicinity of 1\={e} and 3\={e} defects corresponding to (\ref{eq:effH_pm_d}) is plotted in Fig.~\ref{f-defect-th}. The orientation of dumbbells is shown schematically, making it more visible by exaggerating separations between guiding centers. For 1\={e} defects the energy minimum is at $\varphi_i=0$, with the dipole directed towards the defect, and a 2D hedgehog texture results. For 3\={e} defect the energy minimum is at $\varphi_i=\pi/2$ and $3\pi/2$, corresponding to vortices and antivortices with two complex conjugated values of variational parameter $\xi$. The calculation shows that the electric dipole moment for vortices and antivortices  is perpendicular to $\bf{a}_i$ and is directed away from the 3\={e} defect. Thus, the electric dipole moments for vortices and hedgehogs are collinear and oppositely oriented.

In the analytical approach so far, we have included displacements of dumbbells screening the defects away from lattice sites and shape re-arrangements as two independent steps. To check if double-counting in energy decrease is sizable, we performed numerical minimization of the full Hartree-Fock Hamiltonian, in which defect bubbles are introduced and dumbbells are projected onto state (\ref{eq:wf_2e_bubble_gen}). This simulation confirms an appearance of hedgehogs and vortices in the presence of 1\={e} and 3\={e} defects, respectively. Activation energies of defects in the presence of textures computed by analytical minimization of (\ref{eq:effH_pm_d}) and in full numerical simulation are shown by solid and dotted lines in Fig.~\ref{f-defect-th}, respectively.  Energies of topological defects are an order of magnitude smaller than excitation energies for charge defects in the absence of textures, and agree with energy scale observed in experiments.

\begin{figure}
\def\ffile{f-U3e}
\includegraphics[width=0.45\textwidth]{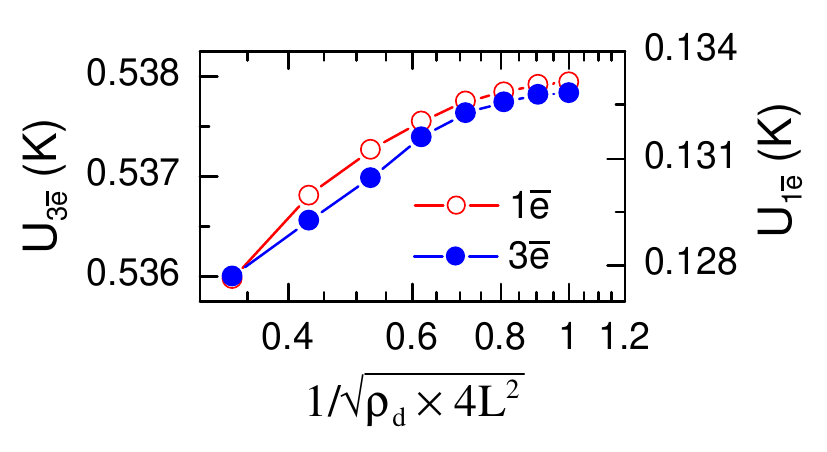}
\caption{Numerically calculated energy of defect-defect interaction vs defect separation shows a crossover from constant to logarithmic dependence at $\approx2L$.}
\label{\ffile}
\end{figure}

 An important result of energy minimization in our system  is that single texture is extended over a finite distance $L\sim 10.5w$ from the charge defect. As distance $R_{ik}$ from the center of the charge defect increases, separation of electrons in a dumbbell $a_i$ decreases. At $R_{ik}\ge L$  round bubbles ($a_i=0$) become energetically favorable. The analytically obtained dumbbell size agrees with the results of numerical simulations, where a crossover from logarithmic dependence of energy per defect on defect separation to almost separation-independent interaction energy is observed at $2L\approx 21 $ for $\nu^*=0.22$, see Fig.~\ref{f-U3e}. In the numerical simulation, $a_i$  on the boundary of defects steeply increases at $R_{ik}< L$. Because of the finite size of textures, displacements of dumbbells do not fully screen the Coulomb potential of the charge defects.
At low defects density, $\rho_d < 1/(4 L^2)$, charged defects interact via such residual Coulomb interaction. At a given density, the energetically favored arrangement is a superlattice superimposed on the bubble crystal, similar to the Wigner \cite{Wigner1934} or Abrikosov\cite{Abrikosov1957} lattice. At a given temperature the equilibrium defect density is achieved via charges coming from and leaving to the contacts. Other channels of equilibration are very slow; for defects separated by distance $L=20w$, a calculated barrier for annihilation of hedgehog defects in the range of $\nu$ where they dominate is $\sim 10$K, for vortex defects a similar calculation gives $\sim 15$K, and for the recombination of vortex and hedgehog for $\nu$, where their excitation energies are equal, the barrier is  $\sim 0.5$K. For two close vortices (or 2D hedgehogs) the barrier is mostly due to the residual Coulomb repulsion described above. Hence it is large for small separations between textures. For close defects of opposite charge the barrier is defined by the balance between attractive force screened by displacements of dumbbels and repulsion due to interactions of dumbbells of two different textures coming into contact.

When $T$ increases or $\nu$ is shifted away from the RIQHE center, the density of defects increases, see Fig.~\ref{f-defect-th}b. When textures overlap at $\rho_d > 1/(4 L^2)$, the last term of energy (\ref{eq:effH_pm_d}) describes the XY-model interaction between dumbbells. Estimating this energy from our analytical model, we take the lower bound on the ``exchange constant'' $J$, given by ${\boldsymbol{\mu}}_i \cdot {\boldsymbol{\mu}}_j$ term in (\ref{eq:effH_pm_d}). Assuming $J$ is defined by the magnitude of elongations $a\sim\lambda$ in the overlap region, we get $J\sim e^2 a^2/\kappa L^3$. This value is in good agreement with the slope of the logarithmic part of the numerically obtained curve Fig~\ref{f-U3e}.
The free energy of topological excitations  is given by\cite{ChaikinLubensky}
\begin{equation}
\label{BKTlog}
E=(\pi J-2T)\ln(\mathcal{L}/w),
\end{equation}
where $\mathcal{L}$ is the size of the system, and the core of topological excitations is $w$.  Such a system must exhibit the Berezinski-Kosterlitz-Thouless (BKT) transition \cite{Berezinskii1972,Kosterlitz1973,Minnhagen1987} at $T_{BKT}=\pi J/2. \approx 5$ mK.   However, unbinding transition in a RIQHE differs from classical BKT transitions. At low defect densities $\rho<1/(4L^2)$ defects textures do not overlap, XY-model is not relevant, and interactions are not logarithmic. Finite size textures do not overlap, loosely forming a defect crystal due to residual Coulomb repulsion, which is an insulating state. The transition occurs at a critical temperature $T_L\gg T_{BTK}$ where $\rho=1/(4L^2)$, once XY model is operative. This transition constitutes {\it melting of a defect crystal}, resulting in mobile defects. Defects 
move as a result of hopping of electrons between crystalline bubbles (dumbbells) and charged bubble defects.

We plot both $T_L(B)$ calculated from the analytic model and from the numerical simulations over the experimentally measured $R_{xx}(T,B)$ in Fig.~\ref{R-T}c. $T_L(B)$ describes the observed phase diagram of the insulator-to-conductor transition rather well.
We can possibly attribute the difference between phase boundaries in experiment and theory to charged impurities in a heterostructure.
Even an undoped GaAs has a residual acceptor density $\sim10^{14}$ cm$^{-3}$ and a smaller  concentration of residual donors. Our simulation shows that charged impurities within $\sim 3\lambda$ off the quantum well lead to the formation of charge 2\=e complexes. Negatively charged impurities form a 2\=e charged complex with a 1\=e bubble defect, and 2\=e bubbles surrounding bubble defect elongate and form a hedgehog. Positively charged impurities form  a 2\=e charged complex with a 3\=e bubble defect, and 2\=e bubbles surrounding bubble defect elongate and form a vortex.  In contrast to unbound 1\=e or 3\=e defects, these textures are attached  to charged defects and do not participate in transport. However, their presence increases the overall density of topological defects and therefore the observed $T_L$.

We note that explanations of the observed \cite{Deng2012a} $R(T)$ Fig.~\ref{R-T}a,b  at temperatures and magnetic fields below $T_L(B)$ curve, which are based on activated transport with $R\propto\exp(-T_a/T)$, or variable-range hopping \cite{shklovskii84} with $R\propto\exp\left[-(T_{ES}/T)^{1/2}\right]$ contradict the experiment. In the former case $T_a=3.1$ K is inconsistent with high mobility measured at 0.03 K. In the latter case the localization length $L_{loc}=k_B T_{ES}\kappa/e^2\sim70$ nm $\sim w/2$ precludes formation of a bubble crystal and the RIQHE state.

\begin{figure}
\def\ffile{R-T}
\includegraphics[width=0.90\textwidth]{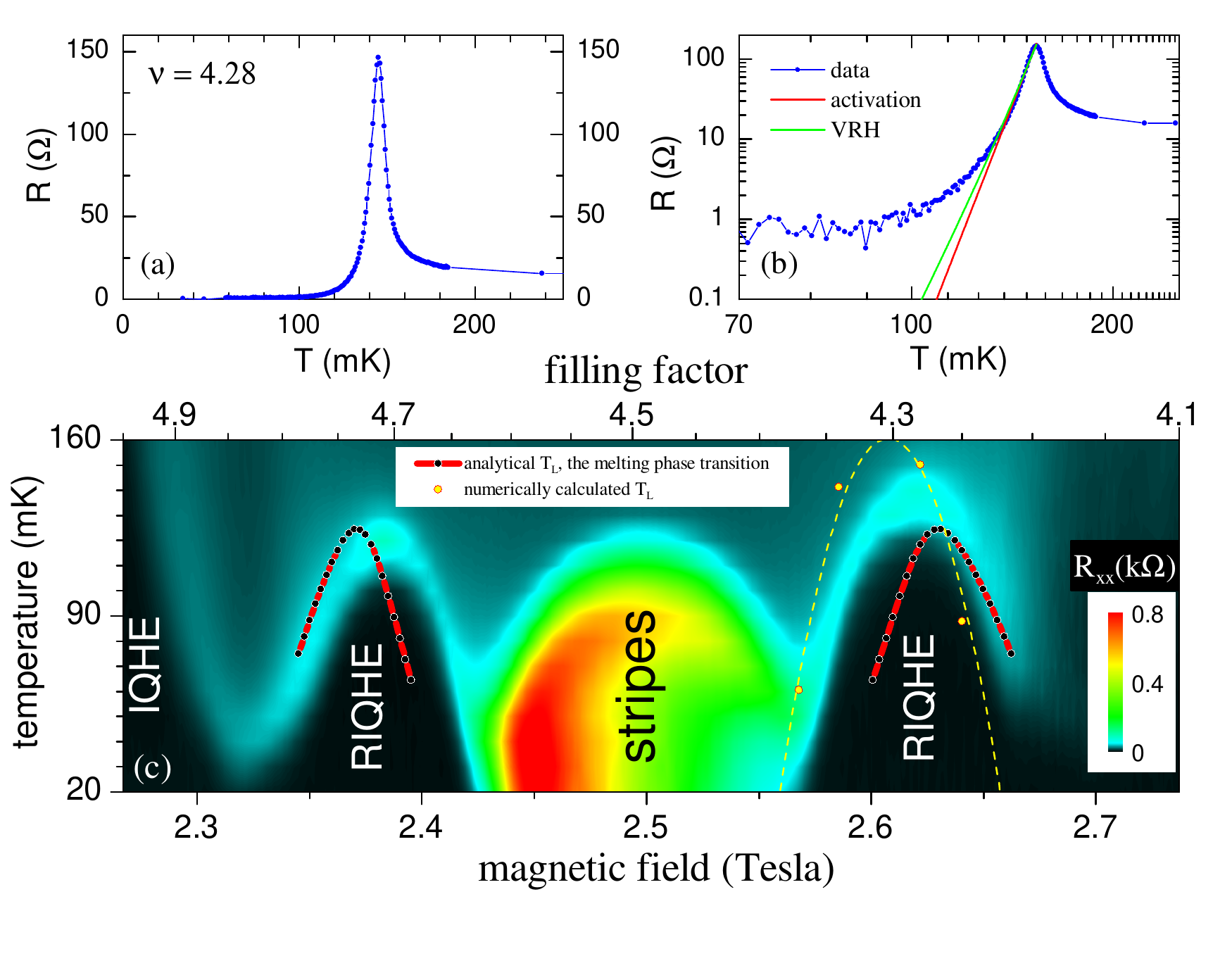}
\caption{(a,b) Temperature dependence of longitudinal resistance in the middle of the RIQHE phase. Fits to the $T$-dependence with activation and variable range hopping give unrealistic $E_a=3.1$ K and $L_{loc}=70$ nm. (c) Analytically and numerically calculated phase boundary $T_L(B)$ is plotted on top of the experimentally measured temperature dependence of $R_{xx}$ and coincides with the sharp increase of conductance at the boundary of isolating and conducting phases.}
\label{\ffile}
\end{figure}

The bubble phase with topological defects persists above $T_L$. The estimate of melting temperature of a bubble crystal due to dislocations \cite{Thouless1978,Fisher1979} is in the range $T_m\approx 250-400$ mK $>T_L$. Therefore in the interval of temperatures between $T_m$ and $T_L$ electrons remain in the bubble phase, but the presence of charge-defect bubbles now permits conduction by tranfer between them and two-electron bubbbles. It is important to note, however, that bubbles are made of electron guiding centers. Thus, most of them are not truly bound. Therefore although this is a correlated phase, conduction can be described by diffusion with somewhat week mobility. Such picture suggests an explanation of the experimental data on the Hall effect obtained here and in \cite{Deng2012a}, showing that when the insulator to metal transition occurs, Hall resistivity experiences rather abrupt transition from RIQHE value to a classical Hall resistivity, which is independent of mobility, like it should be for Hall resistivity in strong magnetic fields. Furthermore, this picture can also explain the longitudinal resistivity. As can be seen from  Fig.~\ref{R-T}, in the temperature range above the transition resistivity and conductivity in the quantum Hall regime decrease when temperature increases. 
We can attribute the decrease in conductivity at temperatures immediately above $T_L$ to the effect of increase in density of dislocations of the bubble phase, as dislocations impede available paths for transer of electrons between bubbles and  charge-defect bubbles. Eventually at $T\approx T_m$ the bubble phase is destroyed, and partially filled LL with electrons or filled LL partially depleted with holes contribute to small background resistivity.  

In summary, we demonstrated that the appearence of charge defects in the bubble crystal corresponding to reentrant integer quantum Hall effect is accompanied by transformation of shapes of the surrounding bubbles to dumbbells.
Depending on the charge of the defect, dumbbells surrounding them form 2D hedgehog or vortex textures. 2D hedgehogs correspond to one-electron defects in 2e bubble crystal and dominate at filling factors closer to the boundary between 2e and 1e bubble crystals. Vortices correspond to three-electron defects in 2e bubble crystal and dominate at filling factors closer to the boundary between 2e and 3e bubble crystals. At suffiuciently high temperatures, textures overlap, interactions of dumbbells is described by XY model, and the defect crystal melts, which explains experimentally observed metal-to insulator transition.
We acknowledge support from the U.S. Department of Energy, Office of Basic Energy Sciences, Division of Materials Sciences and Engineering under Awards DE-SC0010544 (Y.L-G),DE-SC0008630 (L.P.R.),
 DE-SC0006671 (G.C., J.D.W. and M.J.M.).

\pagebreak
\widetext
\setcounter{equation}{0}
\setcounter{table}{0}
\setcounter{page}{1}
\makeatletter
\renewcommand{\theequation}{S\arabic{equation}}
\renewcommand{\thefigure}{S\arabic{figure}}
\renewcommand{\bibnumfmt}[1]{[S#1]}
\renewcommand{\citenumfont}[1]{S#1}
\renewcommand{\thefigure}{S\arabic{figure}}
\renewcommand{\thepage}{sup-\arabic{page}}
\begin{center}
\textbf{\Large Supplementary Materials} \\
\vspace{0.2in} \textsc{Theory of topological excitations and metal-insulator transition in reentrant integer quantum Hall effect}\\
{\it  George Simion, Tzu-ging Lin, John D. Watson, Michael J. Manfra, Gabor A. Cs\'athy, Leonid P. Rokhinson, and Yuli
Lyanda-Geller}
\end{center}
\subsection{Charge Density Wave Phases}
We use Hartree-Fock (HF) method \cite{fogler96,Fogler1997,aleiner95,fukuyama79} to describe charge density wave phases in the $n^{\rm{th}}$ partially filled Landau level\cite{aleiner95}:
\begin{equation}
\label{eq:gen HF_hamilt_S}
H_{\rm{HF}}=\frac{1}{2}\int{\frac{d^2\bf{q}}{(2\pi)^2}
[V_{\rm{H}}({\bf{q})}-V_{\rm{F}}(\bf{q})]|\rho(\bf{q})|^2}~,
\end{equation}
where $\rho(\bf{q})$ is the projected electronic density onto the uppermost LL. Unless otherwise noted, we express distances in units of magnetic length $\lambda$, wavevectors in units of $1/\lambda$, and energies in units of $e^2/\kappa\lambda$, where $\kappa$ is the background dielectric constant.
The Hartree and exchange potentials are:
\begin{eqnarray}
\label{eq:Hartree_pot_S}
V_{\rm{H}}({\bf{q}})&=&\frac{2 \pi}{q}e^{-q^2/2}[L_{n}(q^2/2)]^2~,\\
\label{eq:Fock_pot_S}
V_{\rm{F}}({\bf{q}})&=&2\pi\int{\frac{d^2\bf{q'}}{(2\pi)^2}V_{{\rm{H}}}({\bf{q}})
e^{-i ({\bf{q}}\times {\bf{q'}})\cdot\hat z}}~,
\end{eqnarray}
where $L_{n}$ is the $n^{\rm{th}}$ Laguerre polynomial.

A charge density wave state was proposed as the ground state for 2D systems in the lowest Landau level \cite{fukuyama79} even prior to the discovery of the quantum Hall effect. This prediction appears to be relevant to high Landau levels where different phases compete\cite{fogler-review} and bubble or stripe phases become possible ground states\cite{koulakov96,Fogler1997,Goerbig2003,Cote2003,Ettouhami2006}. In a bubble phase, guiding centers of electron cyclotron orbits form a triangular lattice. A Wigner crystal, a triangular lattice with one electron per lattice cell ($M=1$), is energetically favorable at small effective filling factors $\nu^*=\nu-n_f<0.2$, where $n_f$ is the number of filled Landau levels and $\nu$ is the filling factor. For larger $\nu^*$ bubble phases with $M>1$ can be formed. HF calculations \cite{Ettouhami2006, Goerbig2003, Fogler1997,Cote2003} set a limit $M\le n_f+1$, while density matrix renormalization group method restricts the size of bubbles to $M\le n_f$ \cite{Shibata2001,Yoshioka2002}. A conventional picture of bubble phases is that crystals with $M$ electrons per bubble exist within a certain range of filling factors and the first order phase transitions occur between $M$ and $M\pm 1$ phases. For $\nu^*\approx 0.5$
a stripe phase becomes the ground state.

Considering bubble phases, it is convenient to express the HF Hamiltonian as
a sum of effective interactions between the guiding centers:
$H_{\rm{HF}}=1/2[\sum_{i\neq j}U({\bf R}_{ij})+\sum_i U(0)]$, where $i$ and $j$
labels the nodes of a triangular lattice. An effective interaction $U$ is
given by:
\begin{equation}
\label{eq:Ueff_gen_S} U({\bf{R}}_{ij})=\int{\frac{d^2\bf{q}}{(2\pi)^2}
\rho_{i}^*({\bf {q}}) [V_{\rm{H}}({\bf{q}})-V_{\rm{F}}(\bf{q})]} \rho_{j}({\bf
{q}})\exp{(i\mathbf{q}\cdot{\bf R}_{ij})}~,
\end{equation}
where $\rho_{i}$ represents projected density of a bubble located at the site
$i$. We surmise that HF approach captures physics of the quantum Hall
systems even at low $n$, particularly for the 3rd LL ($n=2$) and 2nd LL ($n=1$)
\cite{Goerbig2003,Cote2003,Ettouhami2006}.

\subsection{Charge Defects and elongations of bubbles in Bubble Crystals}

We now  consider charge excitations of the bubble crystal.
Prior to this work bubbles were almost exclusively treated as entities with uniform charge density. Ettouhami\cite{Ettouhami2004} suggested that guiding centers of electrons in two-electron (2\=e) bubbles are spatially separated even in an ideal bubble crystal with no charge defects. We find that if superposition between wavefunctions of electrons in the same bubble and their phase factors due to magnetic translations are properly taken into account, round shape of bubbles is energetically favorable in an ideal 2\=e bubble crystal. We find, however, that in the vicinity of charged defects, i.e. bubbles lacking one electron (1\=e) or with one extra electron (3\=e), two-electron bubbles become elongated and their shape looks like a dumbbell. The wavefunction of a bubble with two guiding centers separated by $2a$ can be expressed as:
\begin{equation} \label{eq:wf_2e_bubble_gen_S}
\Psi_{\xi,\bf{a}}({\bf{r}}_1,{\bf{r}}_2)=\frac{\psi_{\bf{a}}({\bf{r}}_1,{\bf{r}}_2)
+\xi\psi_{-\bf{a}}({\bf{r}}_1,{\bf{r}}_2)}{\sqrt{2 \left(1-2e^{-2
a^2}a^2\right) \left(1+|\xi |^2\right)+4 \left(1-2 a^2\right) e^{-2 a^2} \Re
e(\xi)}}~,
\end{equation}
where
\begin{equation}
\label{eq:wf_2elebubble_S}
\Psi_{\bf{a}}({\bf{r}}_1,{\bf{r}}_2)=u_{-2}({\bf{r}}_1+{\bf{a}})
u_{-1}({\bf{r}}_2-{\bf{a}})e^{\frac{i}{2}({\bf{r}}_1-{\bf{r}}_2)\times
{\bf{a}}}-u_{-1}({\bf{r}}_1-{\bf{a}})
u_{-2}({\bf{r}}_2+{\bf{a}})e^{\frac{i}{2}({\bf{r}}_2-{\bf{r}}_1)\times
{\bf{a}}}~,
\end{equation}
and $u_{-2}$ and $u_{-1}$ are single electron wavefunctions with angular momenta $m=-2$ and $-1$ in third Landau level ($n=2$) in the symmetric gauge \cite{Giuliani-book}. This is a trial wavefunction similar in spirit to the variational wavefunction proposed by Fogler and Koulakov for round bubbles\cite{Fogler1997}.  The direction of $\mathbf{a}$ characterizes spatial orientation of the dumbbell, and complex parameter $\xi$ permits nontrivial combinations of $\mathbf{a}$ and $-\mathbf{a}$ dumbbell orientations that may potentially emerge in the presence of a magnetic field. Electron density of such a dumbbell, projected on the $n=2$ LL, is
\begin{eqnarray}
\label{eq:density_2el_bubble_S} \rho_{\xi,\bf{a}}({\bf{q}})&=&
e^{-\frac{q^2}{4}} \left[\left(1-2e^{-2 a^2}a^2\right) \left(1+|\xi
|^2\right)+2 \left(1-2 a^2\right) e^{-2 a^2} \Re
e(\xi)\right]^{-1}\nonumber\\
&\times&\left\{e^{i {\bf {q}}\cdot{\bf
{a}}}\left(1+|\xi|^2-\frac{|\xi|^2q^2}{2}\right)+e^{-i {\bf {q}}\cdot{\bf
{a}}}\left(1+|\xi|^2-\frac{q^2}{2}
\right)\right.\nonumber\\
&-&e^{-2a^2+({\bf {q}}\times{\bf {a}})\cdot\hat z} \left[\left(2a^2-({\bf
{q}}\times{\bf {a}})\cdot\hat
z\right)\left(1+|\xi|^2\right)-2\left(1-2a^2\right)\Re e
\xi\right.\nonumber\\
&+&\left.\frac{\xi q^2}{2}-2\xi({\bf {q}}\times{\bf {a}})\cdot\hat
z+i\left(1-|\xi|^2\right){\bf {q}}\cdot{\bf
{a}}\right]\nonumber\\
&-&e^{-2a^2+({\bf {a}}\times{\bf {q}})\cdot\hat z} \left[\left(2a^2-({\bf
{a}}\times{\bf {q}})\cdot\hat
z\right)\left(1+|\xi|^2\right)-2\left(1-2a^2\right)\Re e
\xi\right.\nonumber\\
&+&\left.\left.\frac{\xi^* q^2}{2}-2
\xi^*({\bf {a}}\times{\bf
{q}})\cdot\hat z+i\left(1-|\xi|^2\right){\bf {q}}\cdot{\bf
{a}}\right]\right\}~,
\end{eqnarray}
where $\bar a =a_x+ia_y$, and $a^*=a_x-ia_y$.

We consider 1\=e and 3\=e charge defects, the lowest energy charged excitations of a 2\=e bubble crystal. The wavefunction of a 1\=e defect is $u_{-2}({\bf r})$,  and has a round shape. A 3\=e defect has internal structure with nonuniform density distribution, and its wavefunction is a Slater determinant made of $u_{-2}$, $u_{-1}$, and $u_{0}$. Exact structure and shape of these defects can be determined by minimizing the cohesive energy, however, as our numerical study shows, energetics and electron transport are primarily determined by dumbbells in the vicinity of defects. Thus, the detailed structure of 3\={e} bubbles is not essential and we model it with all three guiding centers being at the same point. Using the dumbbell  wavefunctions (\ref{eq:wf_2e_bubble_gen_S}), expression for charge density (\ref{eq:density_2el_bubble_S}), and wavefunctions for 1\=e and 3\=e defects, we can write an effective Hartree-Fock Hamiltonian for the dumbbell crystal with defects.

\subsection{Numerical simulation of the problem}
\label{numerics}

\begin{figure}
\includegraphics[width=0.95\textwidth]{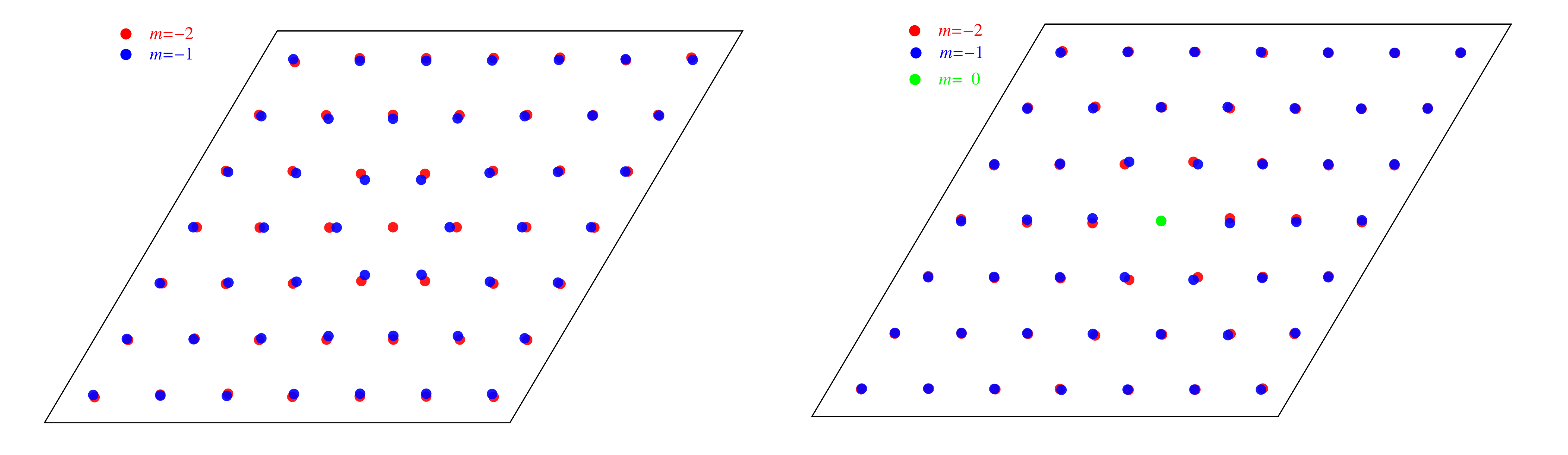}
\caption{{\bf Numerically calculated topological textures surrounding 1\=e and 3\=e defects}. Red and blue circles show positions of electrons with angular momentum $m=-2$ and $m=-1$ correspondingly, green circle is a 3\=e defect.}
\label{NewStructure}
\end{figure}

Full treatment of the effective Hartree-Fock Hamiltonian is an extremely complex endeavor, and we introduce certain simplifications for our numerical simulations. In particular, for $\nu^*$ far away from the center of the 2\=e bubble phase we consider only one type of defects, 1\=e defects at low $\nu^*$ (close to the Wigner crystal) and 3\=e at high $\nu^*$ (close to the phase transition to the 3\=e bubble crystal). We justify this simplification \textit{a posteriori} by demonstrating that the energy to form a 1\=e defect is lower than the energy to form a 3\=e defect at low $\nu^*$ and higher at high $\nu^*$. In order to calculate the minimal energy of the system in the presence of defects we use the following construction. The initial configuration consists of a unit cell with one defect of a given type surrounded by dumbbells placed at the nodes of the bubble crystal. The number of dumbbells in a unit cell is determined by the density of defects. Similar to the treatment of vacancies in a Wigner crystal\cite{Fisher1979} we apply periodic boundary conditions repeating this unit cell, and taking into account interactions off all dumbbells and defects both within the unit cell and between different unit cells. Extra charge on defects cannot be fully screened by elastic displacement of surrounding dumbbells or by transformation of bubbles into dumbbells  (in contrast to screening of vacancies of Wigner crystall at infinity as in \cite{Fisher1979}) because each unit cell has a finite size. Therefore, there is always a residual interaction that makes energetically favorable to position defects as far away from each other as possible, justifying a periodic arrangement of defects. Displacement of dumbbells and magnitude and direction of their elongations are computed iteratively by minimizing total energy (with up to 300 steps), for different defect densities.

Absolute energy minimum corresponds to the bubble crystal with no defects. For finite defect density numerical simulations clearly demonstrate that around 1\=e defects dumbbells emerge forming hedgehog textures, while around 3\=e defects they form vortex textures, Fig.~\ref{NewStructure}. In Fig.~1a of the main text dotted lines show $\nu^*$-dependence of a single defect calculated at a low defect density 0.0025. For comparison, dashed lines show energies of round defects ($a=0$) when only re-adjustments of bubble positions are included. Transformation of 2\=e bubbles into dumbbells lowers defect energies by an order of magnitude. In Fig.2 of the main text we use numerical minimization to plot dependence of energy per defect on defect separation. At high defect densities energy depends logarithmically on defect separation, while at smaller densities energy dependence on inter-defect distance saturates. The crossover is found to be at $\approx 21$ bubble crystal lattice constants. At large separations, 2\=e bubbles close to the boundary of unit cells are positioned at the sites of the ideal bubble crystal and exhibit almost no elongation. Interactions between defects are almost completely (but still not fully) screened by elongated bubbles and, hence, there is almost no energy dependence on defect separation.  Elongations of boundary bubbles steeply increase at defect separations smaller than $21w$, when dependence of energy becomes logarithmic.

\subsection{Analytical model}
In order to gain insight into physics behind these numerical results and to understand the effect of strain on conductivity, we construct an analytical model  based on a small parameter $a/w$, a ratio of a characteristic magnitude of  elongation of 2\=e bubbles, $a\approx\lambda$, to the lattice constant of the bubble crystal $w\approx8\lambda$. Writing an asymptotic expression for the effective interaction of bubbles at different sites of the 2\=e bubble lattice and taking  into account dumbbell shape, it is convenient to express the interaction in terms of elongations $\mathbf{a_i}$ and vectors $\boldsymbol{\mu}_i$, which represent rotated and re-scaled $\mathbf{a_i}$. Retaining terms up to $R_{ij}^{-3}$ for energy change due to re-shaping of bubbles we obtain
\begin{eqnarray}
\label{eq:2elebubble_int_S}
\delta U_2({\bf{R}}_{ij})&=&\frac{({\boldsymbol{\mu}}_i-{\boldsymbol{\mu}}_j)\cdot
 \hat{{ \bf{R}}}_{ij}} {R_{ij}^2}
+\frac{{\boldsymbol{\mu}}_i \cdot
{\boldsymbol{\mu}}_j- 3 ({\boldsymbol{\mu}}_i \cdot \hat{{
\bf{R}}}_{ij})({\boldsymbol{\mu}}_j \cdot \hat{{ \bf{R}}}_{ij})}
{R_{ij}^3}
\nonumber\\
&+&\frac{ u(a_i, \xi_i)+u(a_j,\xi_j)}{R_{ij}^3}+\frac{v(a_i, \xi_i)
\cos(2\phi_i-2\theta_{ij})+ v(a_j, \xi_j)
\cos(2\phi_j-2\theta_{ij})}{R_{ij}^3}~.
\end{eqnarray}
Here $\theta _{ij}$ is the phase of ${\bf{R}}_{ij}$, $\phi _i$ and $\phi_j$ are
the phases of elongation vectors ${\bf{a}}_i$ and ${\bf{a}}_j$ for bubbles
located at sites $i$ and $j$ respectively,
$\mathbf{\hat{R}}_{ij}=\mathbf{R_{ij}}/R_{ij}$, and
\begin{eqnarray}
\label{eq:mu_S}
\boldsymbol\mu&=&\frac{2\cal R({\bf{a}})}{e^{2
a^2}-2a^2} \frac{\sqrt{1-\left(\frac{2 \Re e \xi}
{1+|\xi|^2}\right)^2}} {1+\eta_1(a) \frac{2 \Re e \xi} {1+|\xi|^2}}~,\\
\label{eq:tilde_u_S}
u(a ,\xi)&=& a^2\frac{e^{2 a^2}+2 a^2}
{e^{2a^2}-2a^2} \frac{1+|\xi|^2-2\eta_2(a) \Re e(\xi)} {1+|\xi|^2+2\eta_1(a) \Re e(\xi)}~,\\
\label{eq:tilde_v_S}
v(a,\xi)&=&3 a^2 \frac{e^{2a^2}-2a^2+2} {e^{2a^2}-2a^2}
\frac{1+|\xi|^2+2\eta_3(a) \Re e(\xi)} {1+|\xi|^2+2\eta_1(a) \Re e(\xi)}~,
\end{eqnarray}
where the following notations have been used
\begin{eqnarray}
\label{eq:eta_eff_int_S}\eta_1(a)&= &\frac{1-2a^2}{e^{2a^2}-2a^2}~,\\
\label{eq:eta_tilde_eff_int_S}\eta_2(a)&=&\frac{1-2a^2}{e^{2a^2}+2a^2}~, \\
\label{eq:eta_bar_eff_int_S}\eta_3(a)&=
&\frac{3-2a^2}{e^{2a^2}-2a^2+2}~,
\end{eqnarray}
and $\cal R (\bf {a})$ represents vector $\bf{a}$ rotated by an angle $\phi_{k}=\arg(1-|\xi|^2+2\Im m\xi)$. The above analytical expressions originate from the Hartree term of the potential (\ref{eq:Hartree_pot_S}), contribution from the exchange potential (\ref{eq:Fock_pot_S}) behaves as $\exp{(-R^2/4)}$ and is neglected.

A $a/w$ expansion of the interaction energy (\ref{eq:Ueff_gen_S}) between a 1\=e defect at site $k$ and a dumbbell at site $i$ up to the $R_{ik}^{-3}$ terms:
\begin{equation}
\label{eq:2el1elint_S}
\delta U_1({\bf{R}}_{ik})=\frac{\boldsymbol\mu_i \hat {\bf
{R}}_{ik}}{R_{ik}^2} +\frac{ u (a_i,\xi_{i}) +
v(a_{i},\xi_i)\cos(2\phi_i-2\theta_{ik})}{2 R_{ik}^3}~.
\end{equation}
Similarly, interaction energy between a 3\=e defect and a 2\=e bubble is:
\begin{equation}
\label{eq:2el3el_int_S}
\delta U_3({\bf{R}}_{ik})=\frac{3\boldsymbol \mu_i \hat {\bf
{R}}_{ik}}{R_{ik}^2} +\frac{3\left [ u(a_i,\xi_{i}) + v
(a_{i},\xi_i)\cos(2\phi_i-2\theta_{ik})\right]}{2 R_{ik}^3}~.
\end{equation}
It is worth noting that due to the symmetry of a triangular lattice $\sum_j
\cos(\phi_i-\theta_{ij})/R_{ij}^n=0$, where the summation runs over all sites of
an ideal crystal and $n$ is a positive integer. Also, the following result is
used in what follows: $\alpha=\zeta_1=w^3\sum_{j}1/R_{ij}^3\approx 11.03$.

Effective Hamiltonian of a single charge defect in a dumbbell crystal  is derived using (\ref{eq:Ueff_gen_S}) and (\ref{eq:2elebubble_int_S})-(\ref{eq:2el3el_int_S}), and is given by (3) of the main text:
\begin{eqnarray}
\label{eq:effH_pm_d_S}
H^{\pm}&=&\sum_{i\neq k} \left[\mp\left(\frac{\boldsymbol \mu_i \hat
{\bf{R}}_{ik}}{R_{ik}^2}+ \frac{v_i}{2 R_{ik}^3} \cos{(2\varphi_i)}\right)+
U_f^{\pm}(i))\right]
\nonumber\\
&+&\frac{1}{2}\sum_{i \neq j;~i,j\neq k}\frac{ {\boldsymbol{\mu}}_i \cdot
{\boldsymbol{\mu}}_j- 3 ({\boldsymbol{\mu}}_i \cdot \hat{{
\bf{R}}}_{ij})({\boldsymbol{\mu}}_j \cdot \hat{{ \bf{R}}}_{ij}) }
{R_{ij}^3}~,
\end{eqnarray}
where $(+)$ is for 3\=e and $(-)$ is for 1\=e defects, $\varphi_i=\phi_i-\theta_{i,k}$. The following notations are introduced:
\begin{eqnarray}
v_i&=&v(a_i,\xi_i),\\
U_f^{\pm}(i)&=&U_f^{\pm}(a_i,\xi_i),
\end{eqnarray}
where
 \begin{equation}
\label{eq:en_form_eff_S} U^{\pm}_{f}(a_i,\xi_i)=\left(\frac{\alpha}{w^3}
\mp\frac{1}{2R_{ik}^3}\right) u (a_i,\xi)+\frac{1}{2}U(0,a_i,\xi_i) ~,
\end{equation}
and $U(0,a_i,\xi_i)$ is defined in (\ref{eq:Ueff_gen_S}). The first two terms of (\ref{eq:effH_pm_d_S}) represent an effective one-body (one-bubble) energy, the third term describes a formation energy at the dumbbell site $i$ due to the presence of a defect, and the fourth term represents an effective interaction due to misalignment of dumbbell orientations. In the presence of multiple defects located at sites $k$,
\begin{eqnarray}
\label{eq:effH_pm_dm_S}
H^{\pm}&=\sum_{i\neq k}& \left[\sum_k{\mp\left[\frac{\boldsymbol \mu_i \hat
{\bf {R}}_{ik}}{R_{ik}^2}+ \frac{v(a_i, \xi_i)}{2 R_{ik}^3}
\cos(2\varphi_i)\right]}+ U_f^{\pm}(a_i,\xi_i)\right]
\nonumber\\
&+&\frac{1}{2}\sum_{i \neq j;~i,j\neq k}\frac{{\boldsymbol{\mu}}_i \cdot
{\boldsymbol{\mu}}_j- 3 ({\boldsymbol{\mu}}_i \cdot \hat{{
\bf{R}}}_{ij})({\boldsymbol{\mu}}_j \cdot \hat{{ \bf{R}}}_{ij})}
{R_{ij}^3}~.
\end{eqnarray}

When charged defects are introduced into 2\=e bubble crystal two effects contribute to the lowering of the total energy: transformation of bubbles into dumbbells and displacement of dumbbells from the sites of an ideal bubble crystal . In the numerical solution these two effects are included simultaneously. Finding an analytical solution of our model we also incorporate both effects, but treat them separately. First, we calculate how much energy it costs to create a charge defect in an ideal bubble crystal. Next, we allow displacement of round 2\=e bubbles in order to lower the total energy. Finally, we lower the energy by introducing elongations of 2\=e bubbles, i.e. dumbbells.  While this procedure is approximate, it captures essential physics and allows us to understand numerical results.

\subsubsection{Energy of defects in a perfect triangular crystal in the absence of elongations}

Consider a 2\=e bubble crystal with a microscopic number $N_d$ defects with charge $2-\sigma_d$ ($\sigma_d=\pm 1$). At a fixed filling factor the total number of electrons $2N$ on the top LL is fixed. We consider defects to be many bubble crystal constants $w$ apart.
The ground state is a triangular lattice of $N$ 2\=e bubbles.  When charged defects are present, the total number of bubbles has to change to $N+N_d\sigma_d/2$, as we keep the total number of electrons the same. Assuming all bubbles to be still arranged in a triangular lattice, we find that the change in lattice constant is $\delta w=-w \sigma_d N_d/(4N)$. Then the energy of the lattice with defects is:
\begin{equation}
\frac{1}{2}N\left[1-\left(1-\frac{\sigma_d}{2}\right) \frac{N_d}{N}
\right]\epsilon_2^{(w+\delta w)}+N_d\epsilon_d^{(w+\delta w)}~,
\end{equation}
where $\epsilon_2^{w}$ is the energy required to add one 2\=e bubble into
the bubble crystal with lattice constant $w$,  $\epsilon_d^{w}$ is the energy of a defect embedded in an
otherwise ideal triangular lattice.

From (\ref{eq:Ueff_gen_S}) we obtain that $\epsilon_2^{w}$ is
\begin{equation}
\epsilon_2=\sum_{k=0,1,2}{\zeta_k a_k^{(2)}/w^{2k+1}}+\tilde{U}(0),
\end{equation}
where $\tilde{U}(0)=U(0)/2=833 \sqrt{\pi}/2048$ is the formation energy, $a_0=4$, $a_1=14$, $a_2=765/4$.  Since $w\approx8\lambda$ we restrict asymptotic expansions to three terms. This approach is justified by the comparison with numerical results. Similarly, interaction energy between a defect and 2\=e bubbles is $U_d=\sum_{k=0,1,2}{b_k^{(d)}/r^{2k+1}}$, making $\epsilon_d=\sum_{k=0,1,2}{\zeta_k b_k^{(d)}/w^{2k+1}}+u_d$, where the formation energy of a 1\=e defect is $u_{1}=0$, while for the 3\=e defect $u_3=77463\sqrt{\pi}/65536$. For 1\=e defect $b_0{(1)}=2$, $b_1{(1)}=13/2$, $b_2{(1)}=117/2$, and for 3\=e defect $b_0{(3)}=6$, $b_1{(3)}=45/2$ and $b_2{(3)}=531/2 $. Parameters $\zeta$ are $\zeta_0=-4.2$, $\zeta_1=11.03$ and $\zeta_2=6.76$.

Using expressions for $\epsilon_2$ and $\epsilon_d$ we find the total energy of $N_d$ defects
\begin{equation}
N_d E_d=\frac{1}{2}N\left[1-\left(1-\frac{\sigma_d}{2}\right)
\frac{N_d}{N} \right] \left(\epsilon_2(w)+\frac{\partial
\epsilon_2}{\partial w}\delta
w\right)+N_d\epsilon_d(w)-\frac{1}{2}N\epsilon_2(w)
\end{equation}
and energy per defect:
\begin{equation}
E_d=- \frac{\sigma_d}{8} \frac{\partial \epsilon_2}{\partial w} w+
\frac{\sigma_d-2 }{4}\epsilon_2 +\epsilon_d.
\end{equation}

\subsubsection{Energy of defects in a distorted crystal in the absence of elongations}

We now consider decrease of the total energy when 2\=e bubbles are allowed to adjust their positions while retaining their round shape. We approach this problem in the spirit of Fisher,  Halperin and Morf\cite{Fisher1979}. Due to the presence of charged defects 2\=e bubbles at lattice sites $\mathbf{R_i}$  experience displacements $\mathbf{u}(\mathbf{R_i})$ from their equilibrium  positions. It is assumed that $\mathbf{u}(\mathbf{R_i})\ll w$.  In the framework of elasticity theory, the energy associated with such displacements up to the second order in $\mathbf{u}(\mathbf{R_i})$ is given by
\begin{eqnarray}
E_d(\left\{{\bf{u}}_i\right\})&=&\frac{1}{2}
 \sum_{i,j}{\Pi_{\alpha\beta}({\bf{R}}_i ,{\bf{R}}_j) u_{\alpha} ({\bf{R}_i})u_{\beta }
({\bf{R}}_j)} \nonumber\\
&-&\sum_{i\neq i_0} {\delta V^1_{\alpha} ({\bf{R}}_i)
u_{\alpha}}({\bf{R}}_i) -\sum_{i\neq i_0} {\delta V^2_{\alpha,
\beta}({\bf{R}}_i) u_{\alpha}({\bf{R}}_i) u_{\beta}({\bf{R}}_i)}
\end{eqnarray}
where $\Pi$ is the spring constant matrix,
\begin{equation}
\delta V^1_{\alpha} =\frac {\partial }{\partial r_{\alpha}}
[U_2({\bf{r}})-U_d({\bf{r}})] ~,
\end{equation}
and
\begin{equation}
\delta V^2_{\alpha, \beta} =\frac{1}{2}\frac {\partial^2 }{\partial
r_{\alpha}\partial r_{\alpha}} [U_2({\bf{r}})-U_d({\bf{r}})]~.
\end{equation}
Here the potential energy describing the bubble lattice $U_2$ is given by
\begin{equation}
\label{eq:2elebubble_int_S0}
U_2({\bf{R}}_{ij})=\frac{4}{R_{ij}}
+\frac{14}{R_{ij}^3}+\frac{765}{4R_{ij}^5},
\end{equation}
the interaction energy of 2\=e bubble at site $i$ with a 1\=e defect at site $k$ $U_{d=1}$
is given by
\begin{equation}
\label{eq:1eldef_int_S0}
U_1({\bf{R}}_{ik})=\frac{2}{R_{ik}}+ \frac{13}{2 R_{ik}^3}+ \frac{117}{2 R_{ik}^5} ~,
\end{equation}
the interaction energy of the 2\=e bubble at site $i$ with a 3\=e defect at site k $U_{d=3}$
is given by
\begin{equation}
\label{eq:3eldef_int_S0}
U_3({\bf{R}}_{ik})=\frac{6}{R_{ik}}+\frac{45}{2 R_{ik}^3}+\frac{531}{2 R_{ik}^5}~.
\end{equation}
Multipole terms in expansions (\ref{eq:2elebubble_int_S0}-\ref{eq:3eldef_int_S0}) appear in magnetic field as a result of the shape of the electron wavefunction in the second LL. Because $w$ is large compared to magnetic length, and   $R_{ik}$ is several $w$, we restrict these asymptotic expansions to three terms.

A spring constant matrix is determined in terms of its Fourier-transform, which is related to  the Fourier-transforms of the potentials $\delta V^1_{\alpha}$ and $\delta V^2_{\alpha,\beta}$ by
\begin{equation}
\delta V^2_{\alpha, \beta}({\bf{q}}) =
\frac{1}{2}\sum_{\gamma}V_{\gamma,\gamma}\delta _{\alpha \beta} + A_c
\Pi^d_{\alpha \beta}
\end{equation}
and
\begin{equation}
\delta V^1_{\alpha}({\bf{q}}) = -i A_c\sum_{\gamma}\frac{\partial \Pi^d_{\gamma
\gamma}}{\partial q_a}~,
\end{equation}
where $A_c$ is the area of the elementary cell of the bubble lattice.
Assuming a neutralizing background and writing the potential in the form
\begin{equation}
V({\bf{r}}) =\sum_{k} {\frac {a_k}{r^{2k+1}}}~,
\end{equation}
we obtain an explicit expression for $\Pi$:
\begin{equation}
A_c^2\Pi_{\alpha \beta}({\bf{q}}) =2\pi\frac{q_{\alpha}
q_{\beta}}{q}+\sum_{k} {\frac {\Xi_k a_k}{w^{2k+1}} \left(
q^2\delta_{\alpha\beta} +\frac{4k+6}{2k-1}q_{\alpha}
q_{\beta}\right)}~,
\end{equation}
where constants $\Xi_0=0.26$ , $\Xi_1=2.07$ and $\Xi_2=6.34$.
The change of energy of the lattice in terms of Fourier-transformed quantities is given by
\begin{eqnarray}
E_d(\left\{{\bf{u}}_i\right\})&=& \frac{1}{2} \int{\frac{d
{\bf{q}}}{4\pi^2} {\Pi_{\alpha \beta}({\bf{q}}) u_{\alpha}
({\bf{q}})u_{\beta } ({\bf{q}})}} \nonumber\\
&-&\int{\frac{d {\bf{q}}}{4\pi^2}
{\delta V^1_{\alpha}({\bf{q}}) u_{\alpha}}({\bf{q}})} -\int{\frac{d
{\bf{q}} d {\bf{k}} }{16 \pi^4} {\delta V^2_{\alpha,
\beta}({\bf{q}}-{\bf{k}}) u_{\alpha}({\bf{k}})
u_{\beta}(-{\bf{q}})}}.
\label{eq:Edmin}
\end{eqnarray}
Minimization of this expression assuming
 \begin{equation}
u_{\alpha}({\bf{q}})=\frac{q_{\alpha}}{q^2}f\left(1+cq +dq^2\right)
\end{equation}
yields a decrease in activation energy of the defects caused by displacements $\mathbf{u}(\mathbf{R_i})$.
The obtained energy reduction is very close to the numerical results shown in Fig.~\ref{R-T}a of the main text, where dashed curves show the energy needed to create 1\=e and 3\=e electron defects in the absence of elongations. The corresponding displacements are also close: the bubbles nearest to the defect are displaced by $\sim 0.1w$ in numerical simulation compared with $\sim 0.08w$ from analytical results, and for the displacement of the next nearest neighbors we obtained $\sim 0.03w$ numerically vs $\sim 0.02w$ analytically.

\subsubsection{Lowering defect energy due to re-shaping}

In our analytical approach we minimise elongation-dependent energy (\ref{eq:effH_pm_d_S}) and combine it with the result of displacement minimization (\ref{eq:Edmin}). Minimization (\ref{eq:effH_pm_d_S}) provides the values of the separation vector $a_i$ and the
mixing parameter $\xi_i$. Assuming that $a/R\ll 1$, minimization of the first term in (\ref{eq:effH_pm_d_S}) provides the zeroth order result.  For 1\={e} defects the energy minimum is at $\varphi_i=0$, with the dipole directed towards the defect, and a 2D hedgehog texture appearing. For 3\={e} defect the energy minimum is at $\varphi_i=\pi/2$ and $3\pi/2$, corresponding to vortices and antivortices with two complex conjugated values of variational parameter $\xi$. The calcultion shows that the electric dipole moment for vortices and antivortices  is perpendicular to $\bf{a}_i$ and is directed away from the 3\={e} defect.

 We evaluate the effect of the interaction term by approximating a
dumbbell located far away from the defect as being surrounded by nearest dumbbells with the
same parameters $a$ and $\xi$. The result of such minimization procedure is
\begin{eqnarray}
\label{eq:a_approx} a_i&=&\frac{w^2} {2^{\frac{1}{6}}
3^{\frac{1}{2}}
\alpha^{\frac{3}{2}}}\left(\frac{1}{R_{ik}}\right)^{\frac{4}{3}}-\frac{2^{7/6}w^4}{3^{5/2}
\alpha^{\frac{4}{3}}}\left(\frac{1}{R_{ik}}\right)^{\frac{8}{3}}\\
\label{eq:ci_approx}
\xi_i^{3\bar{e}}&=&-1+i\frac{2 w^2}{2
\alpha^{\frac{2}{3}}} \left(\frac{1}{r}\right)^{4/3}\\
\xi_i^{1\bar{e}}&=&-1+\frac{2 w^2}{2 \alpha^{\frac{2}{3}}}
\left(\frac{1}{r}\right)^{4/3},
\end{eqnarray}
and corresponding activation energies are plotted in Fig.~1a of the main text (solid lines). Although these  energies quantitatively differ from those obtained numerically (dotted lines), qualitatively both approaches (i) result in the decrease of activation energy by an order of magnitude compared to the case of round bubbles (dashed lines), and (ii) predict hedgehog textures around 1\=e defects and vortex texture around 3\=e defects.

An important insight obtained from the analytical model is that textures associated with an isolated defect are extended over a finite distance $L\sim 10.5w$ from the charge defect. As
distance $R_{ik}$ from the center of the charge defect increases, bubble
elongation $a_i$ decreases. At $R_{ik}\sim L$ elongations $a_i=0$ (round bubbles)
become energetically favorable. This happens for defect density that corresponds to a change in the dependence of activation energy on the defect separation in our numerical simulation shown in Fig.~2 of the main text.  We now discuss the significance of these findings.

\subsubsection{Defect crystal, melting and insulator to metal transition}

At low defect density $\rho_d<4 L^2$ textures from different defects do not overlap. At a given $T$ and $\nu^*$ an equilibrium density of defects is established by electrons coming from and leaving to Ohmic contacts, a process accompanied by the formation or destruction of charge defects. Other channels of equilibration, such as annihilation of defects, are very slow.  For defects separated by distance $L=20w$, a calculated barrier for annihilation of hedgehog defects in the range of $\nu$ where they dominate is $\sim 10$K, for vortex defects a similar calculation gives $\sim 15$K, and for the recombination of vortex and hedgehog for $\nu$, where their excitation energies are equal, the barrier is  $\sim 0.5$K. For two close vortices (or 2D hedgehogs) the barrier is mostly due to the residual Coulomb repulsion described above and hence large for small separations between textures. For close defects of opposite charge the barrier is defined by the balance between attractive force screened by displacements of bubbles and repulsion due to interactions of elongated bubbles of two different textures coming into contact. The obtained energy barriers  render annihilation processes ineffective at experimentally relevant temperatures $T<0.15$ K.

The density equilibrium density of defects at small density is given by activational dependence.For a given temperature, there is equilibrium density of defects and the corresponding defect crystal.

Due to residual Coulomb interactions, defects form a \textit{superlattice superimposed on the dumbbell crystal}, somewhat similar to the Wigner\cite{Wigner1934} or Abrikosov\cite{Abrikosov1957} lattice, an arrangement confirmed by numerical simulations
When defect separation becomes $<2L$ textures of neighboring defects start to overlap. It is important to realize that in our analytical model the two-body interaction (\ref{eq:effH_pm_d_S}) represents an XY-model \cite{ChaikinLubensky}. This is transparent if we take a continuum limit $\varphi_i\rightarrow\varphi_j$, where only $\cos{(\varphi_i-\varphi_j)}$ term is important. For the XY-model the vortex and hedgehog textures, which minimize the $1/R^2$ interaction of the bubble system with defects, constitute topological excitations. We note that XY-model physics characterizes dipole-dipole interactions of (\ref{eq:effH_pm_d_S}) even if continuum limit is not applied \cite{Prakash1990}. Thus, for $\rho_d>4 L^2$ energy (\ref{eq:effH_pm_d_S}) includes interaction between dumbbells located near different defects, described by the XY model. Proceeding within the framework of our analytical model and calculating energy caused by such interaction, we take the lower bound on the "exchange constant" $J$, which is the ${\boldsymbol{\mu}}_i \cdot {\boldsymbol{\mu}}_j$ term in (\ref{eq:effH_pm_d_S}). We assume that $J$ is defined by a characteristic magnitude of elongations $a\sim\lambda$ in the region where topological excitations overlap, which sets the lower bound to $J\sim e^2 a^2/\kappa L^3$. Then, as in the XY-model, energy is logarithmic,
\begin{equation}
E=\pi J\ln(\mathcal{L}/w),
\end{equation}
where $\mathcal{L}$ is the size of the system and the core of topological excitations is taken to be of the order of the bubble crystal lattice constant $w$. It is this logarithmical dependence that arises in numerical simulations, Fig.~2 of the main text. Thermodynamical properties of the system are defined by the free energy, which includes the entropy of topological excitations and is given by\cite{ChaikinLubensky}
\begin{equation}
\label{BKTlog-s}
F=(\pi J-2T)\ln(\mathcal{L}/w).
\end{equation}
Such system must exhibit the Berezinski-Kosterlitz-Thouless (BKT) transition \cite{Berezinskii1972,Kosterlitz1973} at $T_{KT}=\pi J/2$. However, the thermodynamic transition in the RIQHE regime differs from conventional BTK transitions, e.g., discussed in \cite{Minnhagen1987}, due to the finite size of topological defects. For $\rho_d<4 L^2$ there is no overlap between neighboring defects (their textures), XY model is not  relavant, and thus, the system cannot undergo the BKT transition. Temperature $T_L$, at which textures from neighboring defects begin to overlap ($\rho_d=4 L^2$) is much higher than $T_{KT}\approx 5$ mK, estimated analytically or extracted from the slope of numerically obtained curves in Fig.~2 of the main text. Thus, $T_L\gg T_{KT}$ in both analytical and numerical calculations and $T_{KT}$ itself is not observed in our experiments. The unbinding of topological defects at $T_L$ required to avoid the divergence of logarithmic interactions (\ref{BKTlog-s}) constitutes {\it melting of the defect rather than the bubble crystal}. The $T_L$'s at different filling factors obtained in our calculations are rather close to the experimentally observed temperatures of metal-insulator transitions.

\subsubsection{The role of residual charge impurities on the metal-insulator transition temperature}

The difference between analytically and numerically obtained $T_L$ are primarily due to the simplified energy minimization procedure in the analytical model. Quite remarkable, however, is about 30\% difference between numerically calculated $T_L$ and experimentally measured metal-insulator transition temperatures. We attribute this difference to the role of disorder not accounted for in our models.We will devote a separate paper to simulations of disorder, but will now briefly sketch some of our results.

The 2D electron gas is separated from the donor layers by symmetric spacer layers of approximately 100nm. Ionized impurities in the doping layer produce a smooth random potential in the quantum well. For large spacers this potential has little effect on the bubble crystal: our numerical simulations show that the ground state of the quantum Hall liquid in the range of filling factors corresponding to 2\=e bubble crystal in the presence of ionized impurities located further than $4\lambda$ from the 2D gas is still a (slightly deformed) 2\=e bubble crystal.

However, even undoped GaAs has a residual acceptor density $\sim10^{14}$ cm$^{-3}$ and a smaller  concentration of residual donors.
Numerical simulations show that depending on the filling factor and on separation from the 2d layer, certain charged impurities within $\sim 3\lambda$ of the quantum well lead to formation of charge 2\=e complexes. Negatively charged impurities form a 2\=e charged complex with a 1\=e bubble defect, and 2\=e bubbles surrounding bubble defect elongate and form a hedgehog. Positively charged impurities form  a 2\=e charged complex with a 3\=e bubble defect, and 2\=e bubbles surrounding bubble defect elongate and form vortex.  In contrast to unbound 1\=e or 3\=e defects , these complexes are strongly localized by charged defects and do not participate in the transport. However, the presence of these complexes increases the overall density of topological defects, lowers the density of the unbound defects needed for melting transition discussed above, and the observed $T_L$.

\subsubsection{Temperature range of free topological defects}

After melting of the defect crystal, topological defects determine the resistivity as long as the bubble crystal is viable. The bubble crystal itself is going to melt, e.g., due to dislocations. We estimate the melting temperature of the bubble crystal using the Thouless formula \cite{Thouless1978,Fisher1979}:
\begin{equation}
\Gamma=\frac{4 e^2 \sqrt{\text{$\pi $n}_s}}{T_m \kappa}=\frac{4
\sqrt{2} e^2 \sqrt{\pi }}{\sqrt[4]{3} T_m w
\kappa},
\end{equation}
where $n_s$ is the bubble density, $w$ is the Wigner lattice constant, $e$ is
the electron charge, $\kappa$ is the dielectric constant, $T_m$ is the melting temperature.
 The factor 4 comes from
the 2\=e charge of the bubbles. The dimensionless
parameter $\Gamma=78-130$ according to
\cite{Thouless1978,Hockney1975,Gann1979,Morf1979}. The estimated $T_m$ can be as low as $\sim 250$ mK, which is above  $> T_L$.


%

\end{document}